\documentclass[prc,twocolumn,showpacs,preprintnumbers,amsmath,amssymb,superscriptaddress]{revtex4}
\date{\today}

\usepackage{graphicx} 
\usepackage{dcolumn}  
\usepackage{bm}       

\begin{document}

\title{Systematic Study of Fission Barriers of Excited Superheavy Nuclei}
\author{J.A. Sheikh }
\affiliation{Department of Physics and Astronomy, University of
Tennessee Knoxville, TN 37996, USA} \affiliation{Physics Division,
Oak Ridge National Laboratory, P.O. Box 2008, Oak Ridge, TN 37831,
USA}
\author{W. Nazarewicz}
 \affiliation{Department of Physics and
Astronomy, University of Tennessee Knoxville, TN 37996, USA}
\affiliation{Physics Division, Oak Ridge National Laboratory, P.O.
Box 2008, Oak Ridge, TN 37831, USA} \affiliation{Institute of
Theoretical Physics, Warsaw University, ul.Ho\.{z}a 69, PL-00681
Warsaw, Poland}
\author{J.C. Pei}
\affiliation{Department of Physics and
Astronomy, University of Tennessee Knoxville, TN 37996, USA}
\affiliation{Physics Division, Oak Ridge National Laboratory, P.O.
Box 2008, Oak Ridge, TN 37831, USA}
\affiliation{Joint Institute for Heavy Ion Research, Oak Ridge, TN
37831, USA}

\begin{abstract}
A systematic study of fission-barrier dependence on excitation energy  has been performed using the self-consistent
finite-temperature Hartree-Fock+BCS (FT-HF+BCS) formalism with the SkM$^*$ Skyrme energy density functional. The calculations have been carried out for even-even superheavy nuclei with $Z$ ranging 
between 110 and  124.  For an accurate
description of  fission pathways, the effects of triaxial and reflection-asymmetric degrees of freedom have been fully incorporated. 
Our survey  demonstrates that
the  dependence of isentropic fission
barriers on excitation energy changes rapidly with particle number,
pointing to the importance of shell effects even at large excitation
energies characteristic of compound nuclei. The fastest decrease of fission barriers with excitation energy is predicted for deformed nuclei around $N$=164
and spherical nuclei around $N$=184 that are strongly stabilized by ground-state
shell effects.  For nuclei $^{240}$Pu and $^{256}$Fm, which exhibit asymmetric spontaneous fission, our calculations  predict a transition to symmetric fission at high excitation energies due to the thermal quenching of static reflection asymmetric deformations. 
\end{abstract}


\maketitle

\section{Introduction}

The mere  existence
of the heaviest and superheavy nuclei with $Z$$>$104 is primarily determined by  shell effects \cite{[Sob66],[Mye66],[Nil69],[Mol94a],[Kru00b],[Ben01],[Cwi05]}. The ground-state (g.s.) shell corrections also determine fission barriers of those systems \cite{[Cwi92],[Ben98a],[Itk02],[Bur04],[Sta05]} as their liquid-drop fission
barriers are negligible. The  discoveries  of new elements
using the cold- and hot-fusion reactions \cite{[Hof00],[Oga07]} over the last decade  provide us with fundamental  information about the structure of the nucleus and the possible existence of the ``island of stability" at the limit of the nuclear mass and charge.

Since the cross sections for production of superheavy nuclei using combinations of available stable projectiles and targets are exceedingly low,
the major experimental challenge is to find optimal conditions that would lead to the synthesis of the species of interest \cite{[Hof00],[Oga07],[Lov07]}. 
Isotopes of elements with 
$Z$ up to 113 have been produced in cold-fusion reactions
using lead or bismuth targets. In these experiments, 
the compound nucleus (CN) is formed at relatively low excitation energies
$E^{*}$ of $\sim$10-12 MeV.
Recently, using the  beams of  $^{48}$Ca and actinide targets, 
superheavy elements with  $Z$=112-116 and 118 have been synthesized \cite{[Oga07]}.  The compound nuclei  formed in such hot-fusion reactions are more neutron-rich than those produced in cold-fusion experiments, and they are
significantly more excited,  $E^{*}$$\sim$36-40\,MeV.

The  crucial quantity that determines  the synthesis of superheavy elements is the CN survival probability \cite{[Den00],[Itk02],[Lov07],[Swi08]},
which strongly depends on the fission barrier characteristics.
Since shell effects are quenched at high temperatures
(see, e.g., Refs.~\cite{[Has73],[Bra74a],[Ign80],[Mor85a],[Gue88],[Egi20]}), 
the stability of the heaviest and superheavy elements with respect to particle emission and fission is expected to strongly depend on excitation energy.

In the previous paper \cite{[Pei09]}, it was demonstrated
that fission barriers of excited superheavy nuclei vary rapidly with particle number. 
The main objective of the present study is to address this question globally by performing systematic calculations of fission barriers of superheavy nuclei as a function of
excitation energy. Our survey has been carried out within the nuclear 
density functional theory (DFT)  generalized to finite temperatures. Guided by results of Ref.~\cite{[Pei09]}, we assume that the fission process is isentropic 
in character. The effects due to the $E^*$ dependence of  triaxial and reflection asymmetric deformations are quantified and the resulting barrier 
damping parameters are extracted.

We also  investigate  the transition from asymmetric to symmetric
fission with increasing excitation energy.  Experimental studies
\cite{[Wag91]} indicate that there is a systematic  increase in the
symmetric mass yield relative to the asymmetric one with excitation
energy. By calculating the reflection-asymmetric deformations along
static fission pathways, we show that such a transition indeed takes
place in selected nuclei.

The  manuscript
is organized as follows. Section~\ref{HFB-T} briefly summarizes the 
FT-HFB formalism. In particular, the need for an isentropic, rather than 
an isothermal, description of the fission process at finite excitation energy is emphasized. 
The particular realization of the FT-HF+BCS model applied in our work is presented
in Sec.~\ref{Model}.
Excitation-energy dependence of fission pathways for two representative nuclei, $^{240}$Pu and $^{256}$Fm,  is discussed in
Sec.~\ref{Results} together with the results  of our systematic calculations of the excitation-energy dependence of the inner fission barrier of superheavy elements. Our  
survey clearly demonstrates that the damping of the first barrier with $E^*$ exhibits an  appreciable dependence on shell effects. Finally, the summary of 
our work is contained in Sec.~\ref{Summary}.

\section{Finite-temperature HFB approach}\label{HFB-T}

Within the mean-field approach, heated nuclei  can be
self-consistently treated by the finite-temperature 
DFT, either   within
Hartree-Fock (HF)
\cite{[Sau76],[Oko87],[Bar85],[Bon85c]} or, if pairing is considered, in the
Finite-Temperature Hartree-Fock-Bogoliubov (HFB) method
\cite{[Egi20],[Mar03],[Kha04],[Kha07],[Min08a]}.
The equilibrium state of a nucleus at a fixed  temperature $T$ and chemical
potential $\mu$ is obtained from the minimization of the grand canonical potential \cite{[Die81],[Bon85c]}:
\begin{equation}
\Omega  = E - TS - \mu N,  \label{grpot}
\end{equation}
where $E$=${\rm Tr} (\hat D \hat H)$ is the average energy, $S$=$-k {\rm Tr} (\hat D \ln \hat D )$ is the entropy,  $N$=${\rm Tr} (\hat D \hat N)$
is the particle-number, and  the density operator $D$ given by
\begin{equation}
{\hat D}  = { e^{-\beta (\hat H - \mu \hat N)} } / {\rm Tr}\left(e^{-\beta (\hat H - \mu \hat N)}\right), \label{denmat} 
\end{equation}
with $\beta = 1/kT$. In the mean-field approximation, the two-body density operator defined in Eq.~(\ref{denmat})
is replaced by a one-body counterpart. The variation of $\Omega$,
 with respect to density, leads to the  temperature-dependent HFB equations~\cite{[Goo81]}:
\begin{equation}\label{hfb}
{\cal H}\left( \begin{array}{c} U_i\\V_i \end{array}\right)=E_i \left(
\begin{array}{c} U_i\\V_i \end{array}\right),
\end{equation}
where ${\cal H}$  is the temperature-dependent HFB Hamiltonian.
Finite-temperature particle and pairing density  matrices \cite{[Mar03]} in the FT-HFB formalism are given by
\begin{eqnarray}
\rho(\beta) &=& UfU^{\dagger} + V^{\ast}(1-f)\tilde V, \\
\kappa(\beta) &=& UfV^{\dagger} + V^{\ast}(1-f)\tilde U,
\end{eqnarray}
and depend on  the Fermi occupations
$f_i = \left(1+e^{\beta E_i}\right)^{-1}$. 

The isothermal scenario,  sometimes assumed in the context of fission process \cite{[Sau76],[Oko87]}, cannot be correct as the compound nucleus is not
in contact with  a heat bath.  Considering the fission as an adiabatic process,
the isentropic picture seems to be   more appropriate \cite{[Die81],[Fab84a]}.
As discussed in Refs.~\cite{[Die81],[Fab84a],[Pei09]},
the two  descriptions of fission can be operationally  related through
the  thermodynamical
identity $\left(\frac{\partial E}{\partial
Q_{20}}\right)_S=\left(\frac{\partial F}{\partial Q_{20}}\right)_T$, which simply states that the generalized driving force associated with the deformation $Q_{20}$ depends only on the state of the system. This identity, useful in practical calculations,  has recently been verified numerically
in Ref.~\cite{[Pei09]} wherein the importance of self-consistency has been pointed out. 

In this work, we shall follow the isentropic picture. The entropy $S=S(T)$ has
been defined as in \cite{[Pei09]}, i.e., it corresponds to the  free energy minimum
at temperature $T_{g.s.}=T$. This value of $S$ is then kept fixed along
the fission path. In this way, the temperature changes with deformation. In
particular, the temperature of the lowest minimum is always greater than
that of the first barrier, and this difference is crucial for the fission
barrier damping.

\section{The Model}\label{Model}

Barrier heights obtained within the HFB
and HF+BCS approaches are quite similar at low temperatures \cite{[Gir79],[Sta07x]}. Moreover, beyond $kT\sim 0.7$ MeV, the two approaches are identical as the static
pairing  vanishes \cite{[Egi20],[Mar03],[Kha07]}. For that reason, in this study we shall present the FT-HF+BCS results  only.

Our FT-HF+BCS calculations were carried out with the Skyrme SkM$^*$ functional~\cite{[Bar82]} in the particle-hole channel. 
This functional
 has been optimized  at large deformations; hence, it is often used
for fission barrier predictions.  In the pairing channel, we employed the
  density-dependent delta
 interaction in the mixed variant \cite{[Dob02c]}:
\begin{equation}
  V(\mathbf{r}-\mathbf{r'}) = V_{0}\left(1-\rho(\mathbf r)/2\rho_0\right)\delta(\mathbf{r}-\mathbf{r'})\,,
\end{equation}
where $\rho_0=0.16$fm$^{-1}$. The pairing-active space in BCS was assumed to consist of the lowest $Z/N$ proton/neutron HF levels.
The pairing interaction strengths
$V_{0}$ are $-438$ and $-372$ (in MeV\,fm$^3$)  for protons and neutrons, respectively.
They were adjusted to reproduce the experimental  odd-even mass differences in
$^{252}$Fm. 

It is known from numerous studies \cite{[Lar72],[Bur04],[Ben98a],[Sta05]} that the first saddle point is lowered by several MeV by triaxial degrees of freedom  and that beyond the first barrier   reflection-asymmetric deformations may become important. Therefore,
when studying saddle points and fission pathways,
it is imperative to employ a model which is capable of breaking  axial and mirror symmetries simultaneously. For that reason, we employed 
a symmetry-unrestricted DFT solver HFODD \cite{[Dob04c],[Dob09b]} capable
of treating simultaneously all  possible collective degrees of
freedom that might appear on the way to fission. 
In the present work, we adopted the HFODD solver to the
FT-HFB and FT-HF+BCS frameworks along the lines of Sec.~\ref{HFB-T}.

\section{Results and analysis}\label{Results}

\begin{figure}[htb]
 \centerline{\includegraphics[trim=0cm 0cm 0cm
0cm,width=0.35\textwidth,clip]{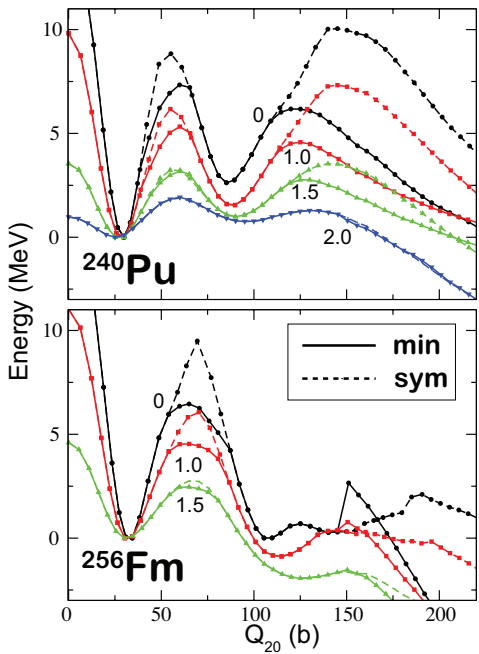}}
\caption{(Color online) 
Fission pathways of $^{240}$Pu (top) and
$^{256}$Fm (bottom) as functions of the mass quadrupole moment $Q_{20}$ at different values of
the ground-state temperature $kT_{g.s.}$ (marked by numbers, in MeV).
 Along the  minimum-energy pathways (``min", solid lines), all self-consistent mean-field symmetries can be broken. To illustrate the corresponding energy gain, the axial, reflection-symmetric energy curves are  also shown (``sym", dashed lines).
The  energy curves have been normalized to zero at the ground-state
minimum. The values of $kT_{g.s.}$= 1, 1.5, and 2 MeV correspond to
excitation energies of 13.82, 36.79, and 70.88 MeV for $^{240}$Pu, and 14.93, 
39.20, and 75.16 MeV (not shown) for $^{256}$Fm.} \label{fig1}
\end{figure}
The main objective of this study is to provide a microscopic description of
fission of excited nuclei, based on the nuclear DFT. To this end, we
solve the constrained  FT-HF+BCS problem along a collective path defined by
a mass quadrupole moment $Q_{20}$. At each value of $Q_{20}$,
self-consistent equations are solved, whereupon the total energy
of the system is always minimized with respect to all remaining  shape
parameters. Along the optimum path found in this way, axial and mirror
symmetries can be broken, i.e., the   multipole moments $Q_{22}$ and/or
$Q_{30}$ may be nonzero. Figure~\ref{fig1} shows the fission pathways
for $^{240}$Pu  and $^{256}$Fm. The former nucleus is known to fission
asymmetrically while the later one is on the edge of the transition from
asymmetric to symmetric fission \cite{[Hul89],[Bro86a]}. It is, therefore,
expected that the fission pathways of these two nuclei would evolve
somewhat differently with increasing excitation energy.

\begin{figure}[htb]
 \centerline{\includegraphics[trim=0cm 0cm 0cm
0cm,width=0.35\textwidth,clip]{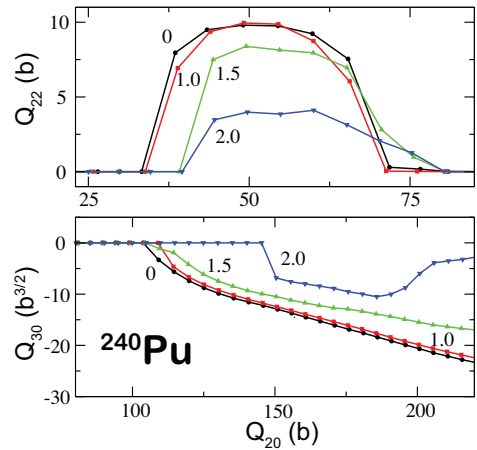}}
\caption{(Color online) Variation of non-axial ($Q_{22}$, top) and reflection asymmetric ($Q_{30}$, bottom) mass moments  as
a function of $Q_{20}$ and temperature  (indicated in MeV)
for $^{240}$Pu. It is seen that triaxiality and reflection asymmetry persist to $kT_{g.s.}$=2\,MeV. However, as indicated in Fig.~\ref{fig1}, their impact on the total energy 
is negligible at the largest temperatures considered.}
\label{fig2}
\end{figure}

For $^{240}$Pu, the optimal fission pathway 
at zero temperature exhibits  the familiar two-humped structure. At
$kT_{g.s.}$=1.0\,MeV  ($E^*$=13.82\,MeV), both saddle points  are reduced by 2-2.5\,MeV.
The isentropic barriers are rapidly quenched with $E^*$, and they become very small at  $kT_{g.s.}$=2\,MeV ($E^*$=70.88\,MeV) due to the thermal melting of shell effects.
In order to assess the impact of  triaxiality on the first, and mirror asymmetry on the second saddle point, we computed the axial reflection-symmetric energy curve for 
$^{240}$Pu (marked as ``sym" in Fig.~\ref{fig1}). The
non-axial ($Q_{22}$) and reflection asymmetric ($Q_{30}$) moments
along the optimal fission pathway are shown in Fig.~\ref{fig2}.
The energy gain on the first barrier
due to  triaxiality, quite appreciable at $T$=0, becomes practically negligible at $kT_{g.s.}$=1.5\,MeV while the corresponding  quadrupole moment $Q_{22}$ is nonzero even at  
$kT_{g.s.}$=2\,MeV. This indicates that at large excitation energies the energy surface of $^{240}$Pu becomes very soft
in the triaxial direction.

A similar conclusion can be drawn for the reflection asymmetric degree of freedom $Q_{30}$
and its impact on the outer barrier.
Experimentally, there is  clear evidence for a transition from
asymmetric to symmetric fission with excitation energy \cite{[Wag91]}.  The results 
displayed in Fig.~\ref{fig1} are consistent with the observed change in the pattern of fission yields. Indeed, at  $kT_{g.s.}$=2~MeV the calculated optimal fission pathway shows
a very weak octupole effect. 

To further explore  the transition from asymmetric to symmetric fission, we now consider $^{256}$Fm. In the heavy Fm isotopes, a sharp transition has been observed \cite{[Hul89]}
from an asymmetric mass division of spontaneous fission products in $^{256}$Fm  to a symmetric mass split in $^{258}$Fm. As seen in Fig.~\ref{fig1}, and discussed in detail
in Ref.~\cite{[Sta07]}, at $T_{g.s.}$=0 the second barrier along the symmetric fission pathway is very broad as compared to the 
asymmetric case, and this explains the asymmetric distribution of fission products observed experimentally.  However, at  $kT_{g.s}$=1.5\,MeV, the symmetric  pathway becomes close in energy   to the asymmetric one. This
indicates that  competition between  asymmetric and symmetric fission is expected to occur in $^{256}$Fm at  lower excitation energies than in $^{240}$Pu.

\begin{figure}[htb]
 \centerline{\includegraphics[trim=0cm 0cm 0cm
0cm,width=0.35\textwidth,clip]{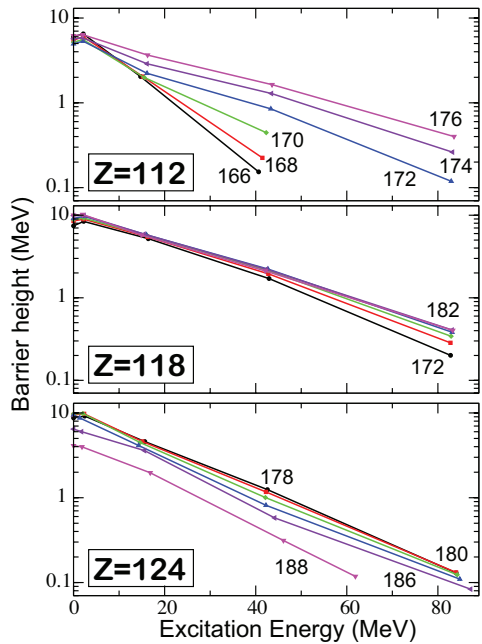}}
\caption{(Color online) Predicted excitation energy dependence 
of barrier heights of even-even superheavy elements with $Z$=112, 118, and 124.} 
\label{fig3}
\end{figure}

We would now like to address the important question  of the synthesis of superheavy elements 
in heavy-ion fusion reactions. It has already been  mentioned that the crucial quanity in the synthesis is the survival probablity, which depends on the
quenching  of the fission barrier height with  $E^*$.
In order to obtain a better understanding of how the shell effects impact the $E^*$ dependence of the first saddle point  of superheavy nuclei,
we performed  systematic FT-HFB calculations  for 48 even-even nuclei 
with 110$\le$$Z$$\le$124 and 
166$\le$$N$$\le$188. A sample result illustrating our methodology is displayed in
Fig.~\ref{fig3} for  $Z$=112, 118, and 124.

The dependence of a fission barrier $(E_B)$ on $E^*$ is usually approximated by a phenomenological expression \cite{[Den00],[Itk02]}
\begin{equation}\label{damp}
E_B\varpropto e^{-\gamma_DE^{*}},
\end{equation}
where the barrier damping parameter
$\gamma_D$ characterizes  the rate of the  barrier quenching with excitation energy.
It is clearly seen from Fig.~\ref{fig3} that the ansatz (\ref{damp}) well describes the FT-HF+BCS results and  the parameter $\gamma_D$ can be 
meaningfully extracted for every nucleus. This is in spite of the fact that many physical effects impact $E_B$-vs-$E^*$
dependence. (In addition to a direct dependence of $E_B$ on entropy, significant contributions come from self-consistent variations 
of nuclear mean fields with $S$, most notably the gradual decrease of 
triaxiality. The quenching of the  pairing energy does not impact the extracted values of $\gamma_D$ as the low-$E^*$ part  of $E_B$ was not considered when extracting the slope of $\ln E_B$.) When inspecting Fig.~\ref{fig3}, one can notice rather dramatic isotonic variations of the damping rate for $Z$=112. 
As discussed in  Ref.~\cite{[Pei09]},
in the isentropic picture, the observed pattern can be attributed to the higher temperature of the lowest 
minimum as compared to that of saddle point. 

\begin{figure}[htb]
 \centerline{\includegraphics[trim=0cm 0cm 0cm
0cm,width=0.45\textwidth,clip]{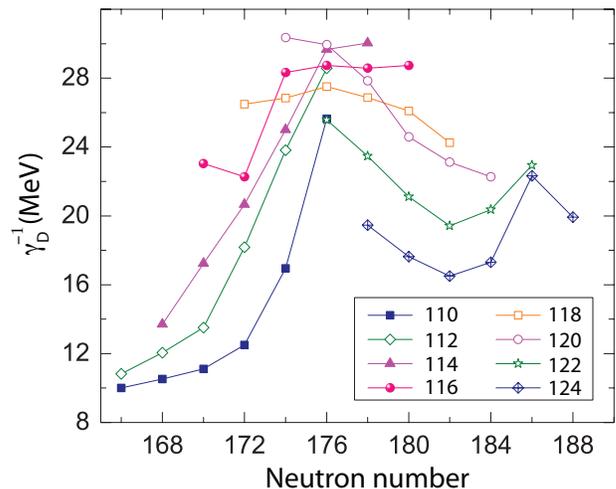}}
\caption{(Color online) Inverse barrier damping parameter $\gamma_D^{-1}$ extracted from
our FT-HF+BCS calculations for 48 even-even superheavy nuclei with 110$\le$$Z$$\le$124 and 
166$\le$$N$$\le$188.} \label{fig4}
\end{figure}
The survey of  $\gamma_D^{-1}$ obtained in this work, shown  in  Fig~\ref{fig4}, nicely illustrates the appreciable  particle number dependence of barrier damping. The maximum of $\gamma_D^{-1}$ is predicted for $N$=176 and 178, while for $N$=166 and 168
$\gamma_D^{-1}$ is fairly small, indicating a rapid decrease of barrier heights with $E^*$ around $^{280}$112, i.e., in the region of deformed superheavy nuclei stabilized by the deformed subshell closure $N$=162 \cite{[Mol94a],[Cwi05]}. For heavier systems
with $Z$=122 and 124, the largest barrier damping effect is expected around $N$=182 and 184, i.e., in the region of the enhanced shell stability around  the expected spherical $N$=184 magic gap \cite{[Kru00b],[Ben01],[Cwi05]}. 
The strong dependence of the barrier damping parameter on $N$ and $Z$ 
indicates the importance of shell effects when modeling 
 the formation of
superheavy elements.  

\section{Summary}\label{Summary}

In conclusion, we performed systematic self-consistent calculations of thermal
fission barriers of superheavy nuclei based on 
the FT-HF+BCS extension of the solver HFODD that is capable of describing arbitrary shapes free from self-consistent symmetry constraints.
Our survey of the fission barrier damping  parameter demonstrates the existence of  strong shell effects on $\gamma_D$. In particular, the fastest decrease of fission barriers with excitation energy is predicted for deformed nuclei around $N$=164
and spherical nuclei around $N$=184 that are strongly stabilized by g.s. shell effects.
On the other hand, for the transitional nuclei around $N$=176, the barrier damping is relatively weak. The particle-number  dependence of $\gamma_D$ shown in Fig.~\ref{fig4} is expected to impact the
survival probability of the superheavy compound nuclei produced
in heavy-ion fusion experiments; we  hope that the values of the  damping parameter  obtained here can be useful in guiding future theoretical work on the production of superheavy nuclei.

We also  studied the quenching of triaxial
and reflection asymmetric deformations with 
excitation energy. For nuclei $^{240}$Pu and $^{256}$Fm, which exhibit asymmetric spontaneous fission, the FT-HF+BCS theory predicts a transition to symmetric fission at higher excitation energies. Finally, the thermal quenching of triaxiality at the first saddle point provides a significant contribution to $\gamma_D$.

Useful discussions with Arthur Kerman, Yuri Oganessian,  and Andrzej Staszczak are
gratefully acknowledged.
This work was supported in part by the National
Nuclear Security Administration under the Stewardship Science Academic
Alliances program through  Grant DE-FG03-03NA00083; by the U.S.
Department of Energy under Contract Nos. DE-FG02-96ER40963 (University
of Tennessee), and DE-AC05-00OR22725 with UT-Battelle, LLC (Oak Ridge
National Laboratory), and DE-FC02-07ER41457 (UNEDF SciDAC
Collaboration). Computational resources were provided by the National
Center for Computational Sciences at Oak Ridge National Laboratory.


\end{document}